\documentclass[prl,twocolumn,showpacs,amsmath,amssymb,superscriptaddress]{revtex4-1}
\usepackage[english]{babel}
\usepackage{amsmath}
\usepackage{graphicx, nicefrac}
\usepackage{float}
\usepackage{siunitx}
\usepackage{lipsum}

\usepackage{dcolumn}
\usepackage{bm,graphicx}
\usepackage{etoolbox}
\usepackage[colorlinks]{hyperref}
\usepackage{url}
\usepackage[normalem]{ulem}
\usepackage[dvipsnames,usenames]{color}
\usepackage[version=3]{mhchem} 

\newcommand{\h}{\hbar}

\newcommand{\kk}{\mathbf{k}}

\newcommand{\e}{\varepsilon}

\newcommand{\om}{\omega}

\newcommand{\al}{\alpha}

\begin{document}

\title{Evidence of 3D Dirac conical bands in TlBiSSe by optical and magneto-optical spectroscopy}

\author{F. Le Mardel\'e}
\author{J. Wyzula}
\affiliation{LNCMI-EMFL, CNRS UPR3228, Univ. Grenoble Alpes, Univ. Toulouse, Univ. Toulouse 3, INSA-T, Grenoble and Toulouse, France}
\affiliation{Department of Physics, University of Fribourg, 1700 Fribourg, Switzerland}

\author{I. Mohelsky}
\affiliation{LNCMI-EMFL, CNRS UPR3228, Univ. Grenoble Alpes, Univ. Toulouse, Univ. Toulouse 3, INSA-T, Grenoble and Toulouse, France}

\author{S. Nasrallah}
\affiliation{Institute of Solid State Physics, TU Wien, A-1040 Vienna, Austria}
\affiliation{Department of Physics, University of Fribourg, 1700 Fribourg, Switzerland}

\author{M. Loh}
\affiliation{Physics Department, Stanford University, Stanford, 94305 California, USA}
\affiliation{Department of Physics, University of Fribourg, 1700 Fribourg, Switzerland}

\author{S. Ben David}
\affiliation{Department of Physics, University of Fribourg, 1700 Fribourg, Switzerland}

\author{O. Toledano}
\affiliation{Departamento Física Interdisciplinar, facultad de Ciencias, Universidad
Nacional de Educación a Distancia (UNED), Avda. Esparta, Las Rozas, 28232, Spain}
\affiliation{Department of Physics, University of Fribourg, 1700 Fribourg, Switzerland}

\author{D. Tolj}
\affiliation{IPHYS, EPFL, Lausanne, Switzerland}

\author{M. Novak}
\affiliation{Department of Physics, Faculty of Science, University of Zagreb, Bijeni\v{c}ka 32, HR-10000 Zagreb, Croatia}

\author{G. Eguchi}
\author{S. Paschen}
\affiliation{Institute of Solid State Physics, TU Wien, Wiedner Hauptstr. 8-10, 1040 Vienna, Austria}

\author{N. Bari\v{s}i\'c} 
\affiliation{Institute of Solid State Physics, TU Wien, A-1040 Vienna, Austria}
\affiliation{Department of Physics, Faculty of Science, University of Zagreb, Bijeni\v{c}ka 32, HR-10000 Zagreb, Croatia}

\author{J. Chen}
\affiliation{Graduate School of Advanced Science and Engineering, Hiroshima University, 1-3-1 Kagamiyama, Higashi-Hiroshima 739-8526, Japan}

\author{A. Kimura}
\affiliation{Graduate School of Advanced Science and Engineering, Hiroshima University, 1-3-1 Kagamiyama, Higashi-Hiroshima 739-8526, Japan}
\affiliation{International Institute for Sustainability with Knotted Chiral Meta Matter (SKCM2), 1-3-1 Kagamiyama, Higashi-Hiroshima 739-8526, Japan}

\author{M. Orlita}
\affiliation{LNCMI-EMFL, CNRS UPR3228, Univ. Grenoble Alpes, Univ. Toulouse, Univ. Toulouse 3, INSA-T, Grenoble and Toulouse, France}
\affiliation{Faculty of Mathematics and Physics, Charles University, Ke Karlovu 5, Prague, 121 16, Czech Republic}

\author{Z.~Rukelj }
\email[]{rukelj@phy.hr}
\affiliation{Department of Physics, Faculty of Science, University of Zagreb, Bijeni\v{c}ka 32, HR-10000 Zagreb, Croatia}

\author{Ana Akrap}
\email[]{ana.akrap@unifr.ch}
\affiliation{Department of Physics, University of Fribourg, 1700 Fribourg, Switzerland}

\author{D. Santos-Cottin}
\email[]{david.santos@unifr.ch} 
\affiliation{Department of Physics, University of Fribourg, 1700 Fribourg, Switzerland}

\date{\today}

\begin{abstract}
TlBiSSe is a rare realization of a 3D semimetal with a conically dispersing band that has an optical response which is well isolated from other contributions in a broad range of photon eneries. We report optical and magneto-optical spectroscopy on this material. When the compound is chemically tuned into a state of the lowest carrier concentration, we find a nearly linear frequency dependence of the optical conductivity below 0.5~eV. Landau level spectroscopy allows us to describe the system with a massive Dirac model, giving a gap $2\Delta =32$~meV and an in-plane velocity parameter $v= 4.0\times 10^5$~m/s. 
Finally, we provide a theoretical recipe to extract all parameters of the anisotropic Dirac band, including the Fermi energy and band degeneracy.
\end{abstract}
\maketitle

%

Research on topological semimetals is an important part of modern condensed matter physics, in great part due to relativistic-like physics which may be observed in such systems \cite{Young2009,AshbyPRB14,Wyzula2021}. An important class of topological semimetals are three-dimensional (3D) Dirac semimetals \cite{Armitage2018,Xiong_2015}.
In them, ideally a single conical band dominates the energy landscape around the Fermi level.
There are not many such systems that we can easily access experimentally \cite{Crassee2018a}.
One such way a conical band can arise is when the bulk energy gap of a topological insulator is gradually closed \cite{YangNatureComm14, Cava2013, Ando2013}, for example by chemical tuning or high pressure \cite{Sato2011, Arakane2012, Xiaoxiang2013}. In such a scenario, the zero-gap state is not topologically protected and a gap may easily open.
In this paper we show that TlBiSSe may be chemically tuned into a clean example of a system whose linear band dispersion dictates the electrodynamic response in a broad energy range.

%
We investigated the low-energy excitations by infrared-spectroscopy and magneto-spectroscopy of TlBiSSe single crystals with an optimized compisition. The single crystal quality of TlBiSSe was improved by using different nominal content of Bi and Tl during synthesis. 
A strong decrease of the metallicity---a reduced Drude contribution--- points to a shift of the Fermi level as the ratio Tl:Bi is tuned. For a specific composition (referred to as sample $S_3$), the system becomes a semimetal or a narrow-gap semiconductor, with the Fermi level close to the Dirac node. The real part of the optical conductivity increases linearly in photon energy up to 0.4~eV, characteristic of a 3D conical band. For the same composition, Landau level (LL) spectra confirm massive Dirac behavior with a band gap as low as $2\Delta = 32$~meV and velocity parameter $v= 4 \times 10^5$~m/s. Our results show that the properties of TlBiSSe depend on the synthesis quality and the ratio of Tl:Bi in the structure, which strongly shifts the Fermi level to expose the large conical band.
 \begin{figure*}[!ht]
	\includegraphics[width=\linewidth]{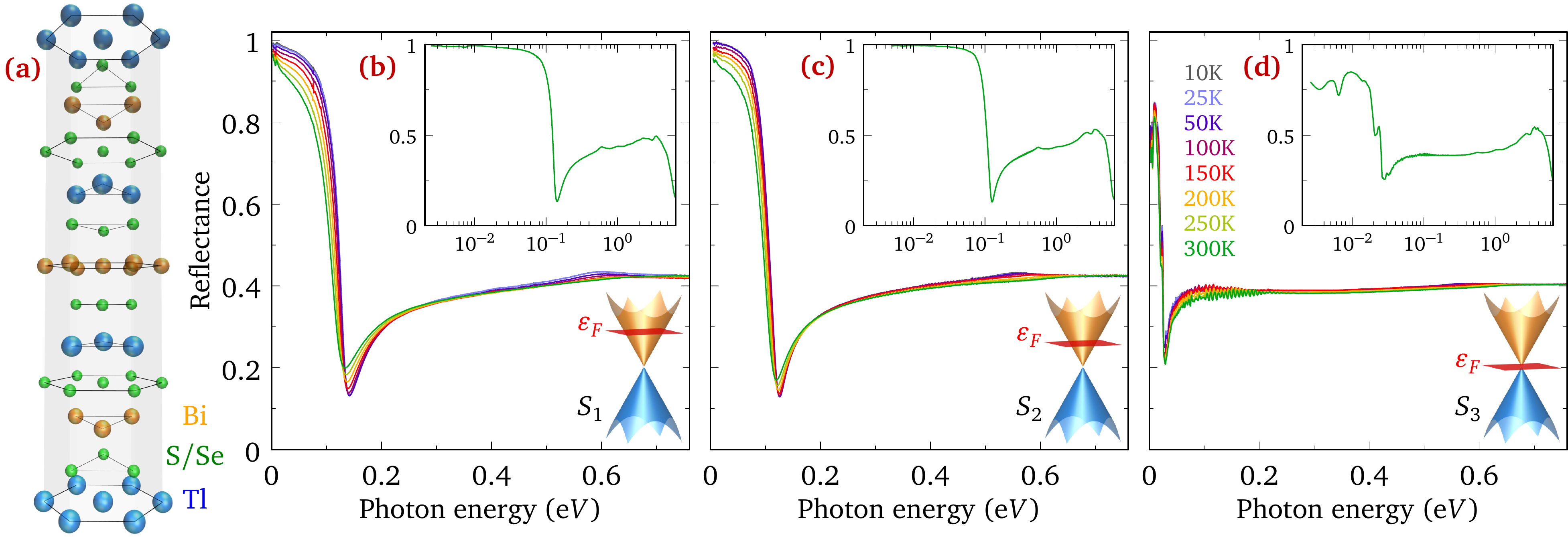}
	\caption{(a) Hexagonal crystal structure of TlBiSSe. (b--d) Reflectance for the samples $S_1$, $S_2$, and $S_3$ up to 0.8~eV for various temperatures. Insets show the reflectance at 300~K in the full energy range up to 6.5~e$V$.}
	\label{fig1}
\end{figure*}
%

We synthesized single crystals of TlBiSSe with a melt-growth technique using high purity elements, 4N or better, of Tl, Bi, S, and Se, sealed under vacuum in a quartz tube. 
The hexagonal crystal structure of TlBiSSe is shown in Fig.~\ref{fig1}(a). 
Recent publications show that stoichiometric melt always gives electron-doped single crystals  \cite{Kuroda2010,Novak2015,Segawa2015} due to bismuth substitution on thallium site. To prevent this, it is possible to play with the initial Tl:Bi ratio to drastically reduce the carrier densities \cite{Novak2015,Kuroda2015}. In this study, we prepared TlBiSSe samples using three different initial ratios of Tl:Bi, starting from Tl:Bi = 1:1 for the synthesis of sample $S_1$, to Tl:Bi = 1.2:0.8 for sample $S_2$, and finally Tl:Bi = 1.5:0.5 for sample $S_3$. We confirmed the high quality crystal structure of these samples by x-ray diffraction analysis. We showed by energy dispersive x-ray diffraction that the final compositions of our grown crystals are close to stoichiometry \cite{SM}.

We measured the near-normal incidence reflectance of freshly cleaved TlBiSSe samples using a Bruker Vertex 70v spectrometer from 10 to 300~K. The sample was mounted on a cold finger and we employed the overfilling technique \cite{Homes1993} to determine its absolute reflectance with an accuracy better than 0.5~$\%$. The data  was extended to 6.4~eV at room temperature using ellipsometry measurements. 
To obtain the complex optical conductivity from a Kramers-Kronig analysis, we extrapolated the low frequency reflectance with a Hagen-Rubens response. At high frequencies, we extended the measurements using the Tanner method \cite{Tanner2015}, which calculates the reflectivity from the atomic x-ray scattering cross sections from 10 to 60 eV, followed by a $1/\omega^4$ free electron behavior. We filled the gap between our experiment and the x-ray data with a smooth cubic spline curve.
The infrared magneto-spectroscopy measurements were performed in a superconducting coil, at 2~K and up to 16~T, while keeping the sample in the He exchange gas. We measured transmission and reflection configurations in the Faraday geometry, where the static magnetic field is parallel to the propagation vector of the incident light wave. The sample measured in transmission was cleaved to a thickness of $\sim 10~\mu$m.
The electronic structure was calculated with the Quantum Espresso package \cite{Giannozzi2009}, employing Projector-Augmented Wave pseudo-potentials to simulate the core interactions \cite{Kresse1999}. We performed calculations in a $2\times2\times2$ trigonal super-cell to recover the Kramers degeneracy and obtain a spin degenerated 3D Dirac cone around the $\Gamma$ point. Then, we unfolded the band structure of this super-cell to a primitive trigonal cell using the BandUp code, more details in the supplementary materials \cite{SM, Medeiros2014,Medeiros2015,Popescu2012}.\\


Figure \ref{fig1} presents the temperature dependent reflectance, as a function of the photon energy, measured for three different batches of TlBiSSe single crystals: $S_1$, $S_2$ and $S_3$. The insets show the full energy range measurements performed at room temperature.
In the low energy range, the reflectance of samples $S_1$ and $S_2$, in panels (b) and (c) respectively, show similar responses which is typical of a  system with a small Fermi surface. The 10~K reflectance tends to unity at very low energy, $R(\omega \rightarrow 0)\rightarrow 1$. The reflectance in that range strongly decreases with the increase of the temperature. 
Around 100~meV, the reflectance of both samples dramatically drops by about 80\% at the screened plasma edge. The sharp plasma edges indicate that the Drude scattering rate in both samples is fairly low.\\
In contrast, the low energy reflectance of sample $S_3$, in panel (d), shows a response more characteristic of an undoped semiconductor or a semimetal, with several strong phonon modes below 30~meV. The intensity of these modes decreases with the increase of temperature. The reflectance for the composition $S_3$ does not approach unity at the lowest attained energies, although it has a weak upturn. The strong edge seen at 20~meV is not related to Drude carriers, but is rather linked to a strong phonon response.
Moreover, in sample $S_3$ we observe clear oscillations in the reflectance between 50 and 200~meV. These are Fabry-Perot interference fringes caused by a transparent energy window in this sample.

  \begin{figure*} [!t]
	\includegraphics[width=\linewidth]{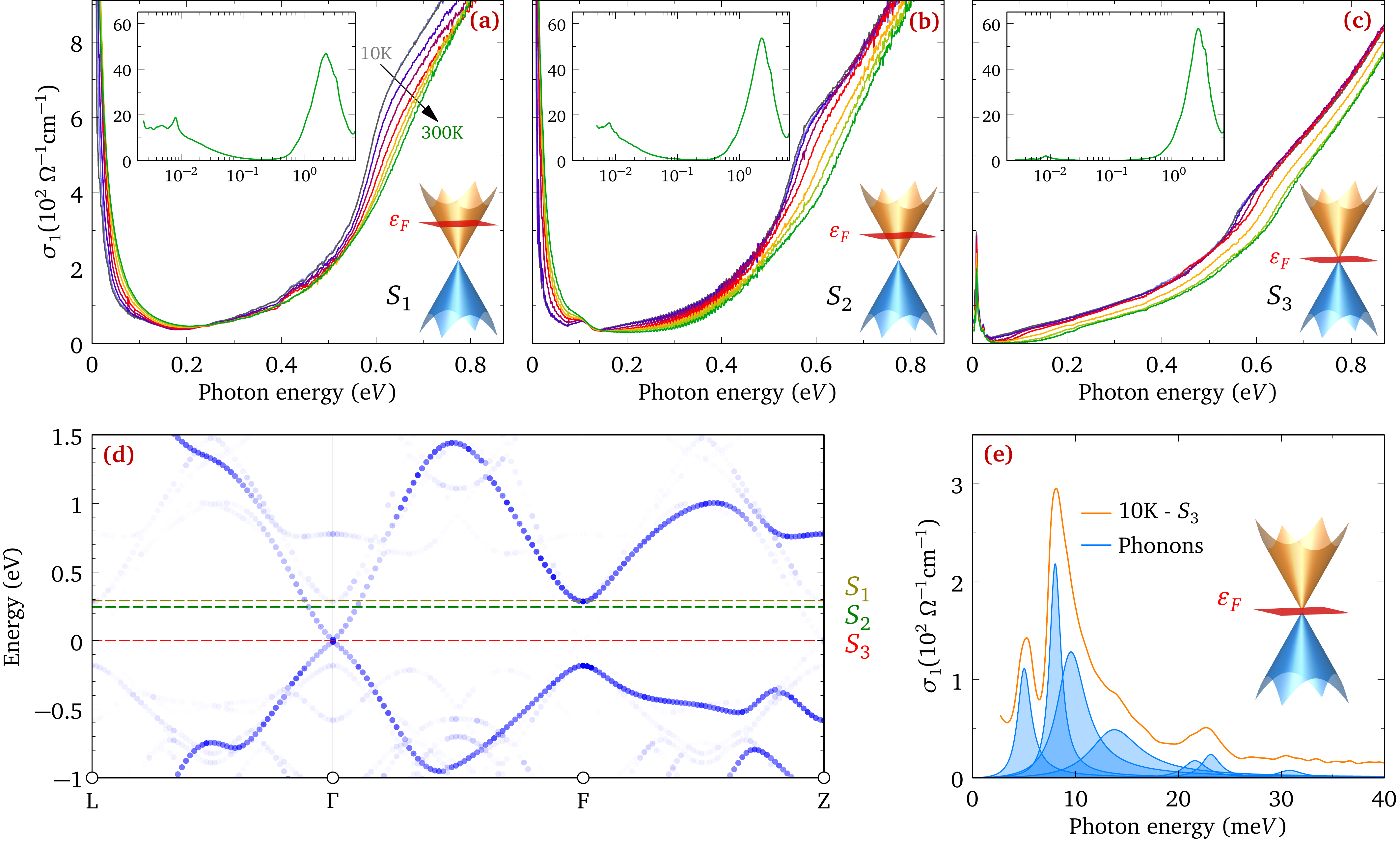}
	\caption{Energy dependence of the real part of the optical conductivity, $\sigma_1(\omega)$, for samples (a) $S_1$, (b) $S_2$, and (c) $S_3$. The main panels show the low energy range, while the insets show the full energy range of $\sigma_1(\omega)$ at 300~K on a logarithmic scale. The optical conductivity is obtained from Kramers-Kronig transformation on the reflectance data shown in Fig.~\ref{fig1}. (d) Band structure calculations in a trigonal primitive cell after unfolding the $2\times2\times2$ trigonal super-cell band structure. The blue color scale represents the amplitude of the super-cell eigenstates projection into the primitive cell eigenstates \cite{SM}. Dashed lines indicate the approximate positions of Fermi level for the three samples. (e) Low energy optical conductivity including a fit of the phonon modes \cite{Kuzmenko2005}.}
	\label{fig2}
\end{figure*}

Overall, the plasma edge shifts a little towards lower energies moving from sample $S_1$ to sample $S_2$, while for sample $S_3$ the free-carrier plasma edge moves outside of our experimental window. This means that the carrier density strongly decreases as we tune the composition from $S_1$ to $S_3$ by increasing the Tl:Bi ratio. This initial observation is in line with the previous transport measurements \cite{Novak2015,Kuroda2015}.
At higher energies, all three samples present a similar reflectance behavior: (1) An almost flat region in the range 250 - 800~meV with a reflectance value of around 0.4, (2) a small bump at 10~K around 550 - 600~meV, weakened at higher temperatures, and (3) a strong decrease of the intensity above 2.5~eV at 300~K.

Figure \ref{fig2}(a--c) shows the energy dependence of the real (dissipative) part of the optical conductivity, $\sigma_1(\omega)$, in the low energy range. Each panel corresponds to a different composition, from $S_1$ to $S_3$. 
At low energies, $\sigma_1(\omega)$ of samples $S_1$ and $S_2$ shows a well-defined Drude peak, which narrows with a decreasing temperature. For the 10~K data, we fit the reflectance of samples $S_1$ and $S_2$ \cite{Kuzmenko2005,SM} and obtain unscreened plasma frequencies of around $540$~meV and $490$~meV, with scattering rates of  $\sim 3.2$~meV and $\sim 2.5$~meV, respectively. For both samples, $\sigma_1(\omega)$ has a poorly conducting intermediate energy region immediately above the Drude component, ending with a sudden onset of absorption around 550 meV.
Comparing with the band structure calculation in Fig.~\ref{fig2}(d), we may assign this absorption onset to transitions near the F point in the Brillouin zone \cite{Singh2012}.
Overlaying the curves of the samples $S_1$ and $S_2$, we notice a relative shift of around 40~meV of the absorption onset which allows us to estimate the position of their Fermi levels. We may place them at $\e_F \sim$ 280~meV above the Dirac node for sample $S_1$, and at $\e_F \leq $  260~meV  for sample $S_2$. 
A rough sketch of the Fermi level position is given in Fig.~\ref{fig2}(d), plotted on top of the DFT-calculated bands.\\

On the other hand, $\sigma_1(\omega)$ of sample $S_3$ does not contain any visible Drude contribution, most likely because it is very narrow and limited to energies which are too low for us to access optically. Instead, several infrared active phonon modes can be seen at approximately 5, 8, 14 and 23~meV, and a possible weak mode at $\sim 30$~meV. All these modes are shown in Fig.~\ref{fig2}(e). We notice that the phonon modes are generally rather broad, which may be caused by disorder in the chemical structure. An obvious source of such disorder comes from a random distribution of S and Se atoms, which occupy the same lattice positions. The DC anisotropy resistivity measurements performed on sample $S_3$ \cite{SM} show bad metal behavior in both the \textit{ab}-plane and c-axis. However, we obtain a very low DC conductivity value of around 310 $\Omega^{-1}$cm$^{-1}$ in the \textit{ab}-plane, one order of magnitude smaller as compared to the value of similar crystals obtained in a previous study \cite{Novak2015}.

The absence of a Drude term in the optical conductivity confirms that increasing the initial Tl:Bi ratio can drastically reduce the carrier densitiy in this system and shift the Fermi level closer to the Dirac node \cite{Novak2015,Kuroda2015}.
When the Fermi level is in a close vicinity of the Dirac node, then the low energy interband transitions become apparent through a linear energy dependence of $\sigma_1(\omega)$ in the range 35 to 320~meV. 
At higher energies, an additional contribution kicks in at around 520~meV, and like in samples $S_1$ and $S_2$, it may be assigned to transitions near the F point of the Brillouin zone.

Finally, the optical conductivity at 300~K in the very high energy region of all three samples is shown in the insets of Fig.~\ref{fig2}(a--c). 
A strong interband transition appears around 2.5~eV. This peak in $\sigma_1(\omega)$ is a fingerprint of a van Hove singularity or a saddle point. It appears because of the Dirac band folding, which can be seen as a plateau in the $\Gamma$--F direction in Fig.~\ref{fig2}(d). Such a saddle point often leads to strong interband transitions and may be seen in the optical conductivity of several Dirac systems \cite{Martino2019,Santos-Cottin2021,Mak2011,Ebad2019,DSC2020,FLM2020}.

\begin{figure*}[!t]
	\includegraphics[width=\linewidth]{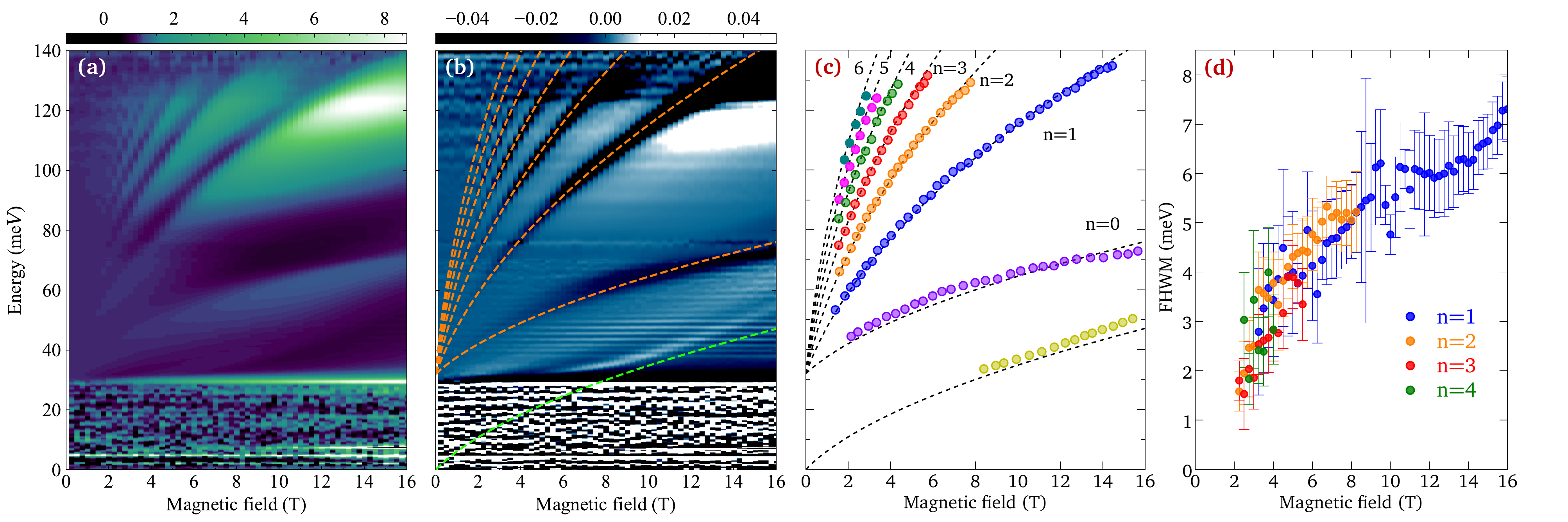}
	\caption{Color plot of (a) the relative magneto-transmission, T$_B$/T$_0$, and (b) its derivative d$/$d$B$ [T$_B$/T$_0$]. 
	(c--d) Magnetic field dependence of the energy positions and FWHM of the LLs fitted using Gaussian distributions.}
	\label{fig3}
\end{figure*}

The band dispersion is visible in zero-field optical conductivity through the density of states (DOS). However, for a system with a very small carrier density, it can be instructive to apply magnetic fields. 
Since the sample $S_3$ has a low reflectance, in the magneto reflection measurements we only observe two broad intra-LLs excitations crossing the phonon modes in the low energy up to 45~meV \cite{SM}.
Figure~\ref{fig3}(a) shows the relative magneto-transmission, the transmission in magnetic field divided by the one at zero magnetic field, $T_B/T_0$ of sample $S_3$ measured in the energy range 30--140~meV, the exact region where we observe the Fabry-Perot interference fringes in the zero field reflectivity measurement discussed above, see Fig. \ref{fig1}(d). Its energy derivative is shown in Fig.~\ref{fig3}(b). A series of inter-LL transitions can be seen in both magneto-transmission and its derivative. 
Those transitions extrapolate to a finite energy in the limit of a vanishing magnetic field, which implies a finite band gap. Moreover, the transitions are sublinear as a function of B. This is a signature of a strongly non-parabolic band dispersion.
We can fit the energies of inter-LL transitions using a massive Dirac model, shown in Fig.\ref{fig3}(b) \cite{SM}. In this model, assuming $k=0$, each LL has an energy given by:
 \begin{equation} \label{LL}
 \varepsilon_n^{\pm}(B) =\pm \sqrt{2\hbar ev_xv_yBn + \Delta^2},
\end{equation}
where $n$ is an integer, $2\Delta$ is the band gap, and the parameter $v_x$ and $v_y$ are in-plane asymptotic velocities of the Dirac cone.

Figures \ref{fig3}(c) and (d) show the energies of the observed inter-LL transitions as a function of B, together with the corresponding full width at half maximum (FWHM) of each transition, extracted using a Gaussian fit. Interestingly, the FWHM of the inter-LL excitations increases monotonously with the magnetic field.
Finally, the dashed lines in Fig.~\ref{fig3}(b) and (c) show the results of the inter and intra--LL transition fit, using a selection rule $n \rightarrow n \pm 1$.
The performed fitting procedures implies the following parameters: $2\Delta\sim 32$ meV and $\sqrt{v_xv_y}=4.0 \times 10^5$~m/s.
This velocity parameter is close to the slope of the conical band measured by photoemission \cite{Souma2012,YangXU2011}, and also comparable to other systems \cite{DSC_TaAs2022, Orlita2014, Crassee2018}.
The slope of this conical band also enters the linear part of $\sigma_1(\omega)$ in Fig.~\ref{fig2}(c). 

For a 3D conical band system, the low-energy $\sigma_1$  at zero-temperature is described by \cite{Rukelj2020,Lim_2020,Kotov2016}:
\begin{equation} \label{sig1}
\sigma_1 (\omega)= g \frac{\sigma_0}{6\pi} \frac{\hbar\omega}{v} \Theta(\hbar\omega-2\e_F),
\end{equation}
where $\sigma_0=e^2/(4\hbar)=6\times 10^5$~$\Omega^{-1}$, $v$ is the velocity parameter related to the slope of the conical bands, and $g$ is the degeneracy of the Dirac cone in the Brillouin zone. It combines the spin and valley degeneracy, and in our case $g=2$ since we have a spin-degenerate Dirac cone at the $\Gamma$ point. The above formula is also approximately valid for the massive Dirac case when the photon energy  $\hbar\omega \sim \e_F  \gg \Delta$, where the bandgap $2\Delta = 32$~meV is obtained from the magneto-optical measurements.\\

However, comparing the result of the formula \ref{sig1} with optical conductivity data in Fig. \ref{fig2}~(c), implies the velocity $1.8 \times 10^5$~m/s, a value significantly lowers as compared to above extracted $\sqrt{v_x v_y}$. One reason for such a disagreement is that $\sigma_1$ probes the joint density of states which reflects the dispersion of electrons in all three spatial directions while photoemission and magneto-optics just in the plane.  

Previous calculations of the electronic band structure \cite{Singh2012, Niu2012} suggest that the out-of-plane velocity, $v_z$, is approximately half the in-plane velocity. Moreover, ARPES measurements performed on TlBi(S$_{1-\delta}$Se$_\delta)_2$ show that the in-plane Fermi surface around the Dirac cone at $\Gamma$ becomes more anisotropic as the content of selenium increases \cite{YangXU2011}.\\
If we take into account such a non-isotropic linear dispersion into our zero field reflection measurements, then $v$ of sample $S_3$ can be approximated by \cite{SM}:
\begin{equation}\label{b2}
    \frac{1}{v}  =  \frac{(v_x^2 + v_y^2)}{2v_x v_y v_z}  \approx \frac{1}{1.8\times 10^5 \, \text{ m/s}}.
\end{equation}
Using the expression (\ref{b2}), the extracted in-plane velocity $\sqrt{v_xv_y}$ from the magneto-optical measurements, and the DC anisotropy ratio $R \sim$ 2.7 shown in Fig.~S5, we extract the three velocities ($v_{x}, v_{y}, v_{z}$) of the Dirac cone \cite{SM}:
\begin{eqnarray}
&& v_x \approx 7.3\times 10^5 \, \text{ m/s}, \nonumber \\
&& v_y \approx 2.2\times 10^5 \, \text{ m/s}, \nonumber \\
&& v_z \approx 3.3\times 10^5 \, \text{ m/s}. 
\end{eqnarray}
Supposing a rigid band shift for the three different dopings $S$, together with the extracted velocities values and parameters obtained from the Drude model, we obtain an approximate Fermi level $\e_F \simeq 290$~meV and $240$~meV for the samples $S_1$ and $S_2$ \cite{SM}. These values are in good agreement with the ones estimated from the onset of absorption in Fig.~\ref{fig2}.
Finally, with the value of the Fermi level and the asymptotic velocities, we estimate, assuming solely the Dirac cone, the total concentration of electrons $n$ for sample $S_1$ and $S_2$ to be $n_1 = 5.6 \times 10^{19}$ cm$^{-3}$ and $n_2 = 2.9 \times 10^{19}$ cm$^{-3}$ \cite{SM}. These values agree well with previously values extracted from the magneto-transport measurements \cite{Novak2015}.\\
 
The lowest inter-LL transition sets in at 2--3~T, see Fig.~\ref{fig3}(a--b), which gives us the quantum limit of this sample composition. This is the field for which all the carriers are confined to the lowest LL. 
In the $S_3$ sample, a considerable chemical disorder is present and reflected, \emph{e.g.}, in the width of the observed phonon lines. In such a situation, it may be surprising that we do observe such well-defined quantization of Landau levels. In fact, the $S_3$ sample has a relatively low electron density, and consequently, the Fermi energy $\e_F$ lies relatively close to the (weakly gapped) Dirac point. In such a case, the corresponding cyclotron mass (defined classically, $m_c=\e_F/v^2$) is small and implies sufficiently large spacing of LLs to overcome the impact of disorder.\\

In conclusion, we have shown that TlBiSSe is indeed a 3D Dirac system with an anisotropic band dispersion. Tuning the Tl:Bi ratio during the growth can bring the Fermi level near the neutrality point. Our zero-field optical spectra show signatures of a broad range linear band dispersion, up to 0.4~eV, leading to the optical conductivity linear in energy, $\sigma_1(\omega) \propto \hbar\omega$. 
The magneto-optical spectra show a series of inter-LL transitions which confirm that the in-plane dispersion of the bands in the lowest-carrier-density sample is nearly linear, with only a small band gap of $2\Delta \sim 32$~meV and an average in-plane velocity of $4.0 \times 10^5$~m/s. 
Therefore, TlBiSSe represents a clean example of a 3D Dirac semimetal which hosts a a single conical band defining the optical and magneto-optical response in a broad range of photon energies. \\


We thank P. Hofegger for the technical support.
A.A. acknowledges funding from the  Swiss National Science Foundation through project PP00P2\_202661.
This research was supported by the NCCR MARVEL, a National Centre of Competence in Research, funded by the Swiss National Science Foundation (grant number 205602).
D.S-C. acknowledges for the SPARK grant CRSK-2\_196610 from the Swiss National Science Foundation.
Z.R was funded by QautiXLie Centre of Excellence (Grant KK.01.1.1.01.0004). 
M.N. and N.B acknowledge support of CeNIKS project co-financed by the Croatian Government and the EU through the European Regional Development Fund Competitiveness and Cohesion Operational Program (Grant No. KK.01.1.1.02.0013).
S.N. and N.B acknowledge the support of the European Research Council (ERC Consolidator Grant No. 725521)
A.K. acknowledges financial support from funding from KAKENHI (Nos. 17H06138, 18H03683).
This work has been supported by the ANR projects DIRAC3D (ANR-17-CE30-0023) and COLECTOR (ANR-19-CE30-0032). We acknowledge the support of LNCMI-CNRS, a member of the European Magnetic Field Laboratory (EMFL).

\onecolumngrid
\section{Appendix}
\subsection{Samples characterization}
We synthesized single crystals of TlBiSSe with a melt-growth technique using high purity precursors (4N or better) of Tl, Bi, S, and Se sealed under a vacuum quartz tube. Recent publications have shown that stoichiometric melt always gives electron-doped single crystals due to bismuth substitution on thallium site. To prevent that, it is possible to play with the initial Tl:Bi ratio which allows a drastic reduction of the carrier densities \citep{Novak2015, Kuroda2015}. In this study, we prepared TlBiSSe samples using three different initial ratios of Tl:Bi: Tl:Bi = 1:1 for Sample $S_1$, Tl:Bi = 1.2:0.8 for Sample $S_2$, and Tl:Bi = 1.5:0.5 for Sample $S_3$. We confirmed the conventional hexagonal crystal structure Fig.~\ref{BZs}(d) of the samples by x-ray diffraction analysis (XRD) and that the compositions of our grown crystals are close to stoichiometry by energy dispersive x-ray analysis (EDX).
\subsubsection{EDX}
\begin{figure*}[!th]
	\includegraphics[width=10cm]{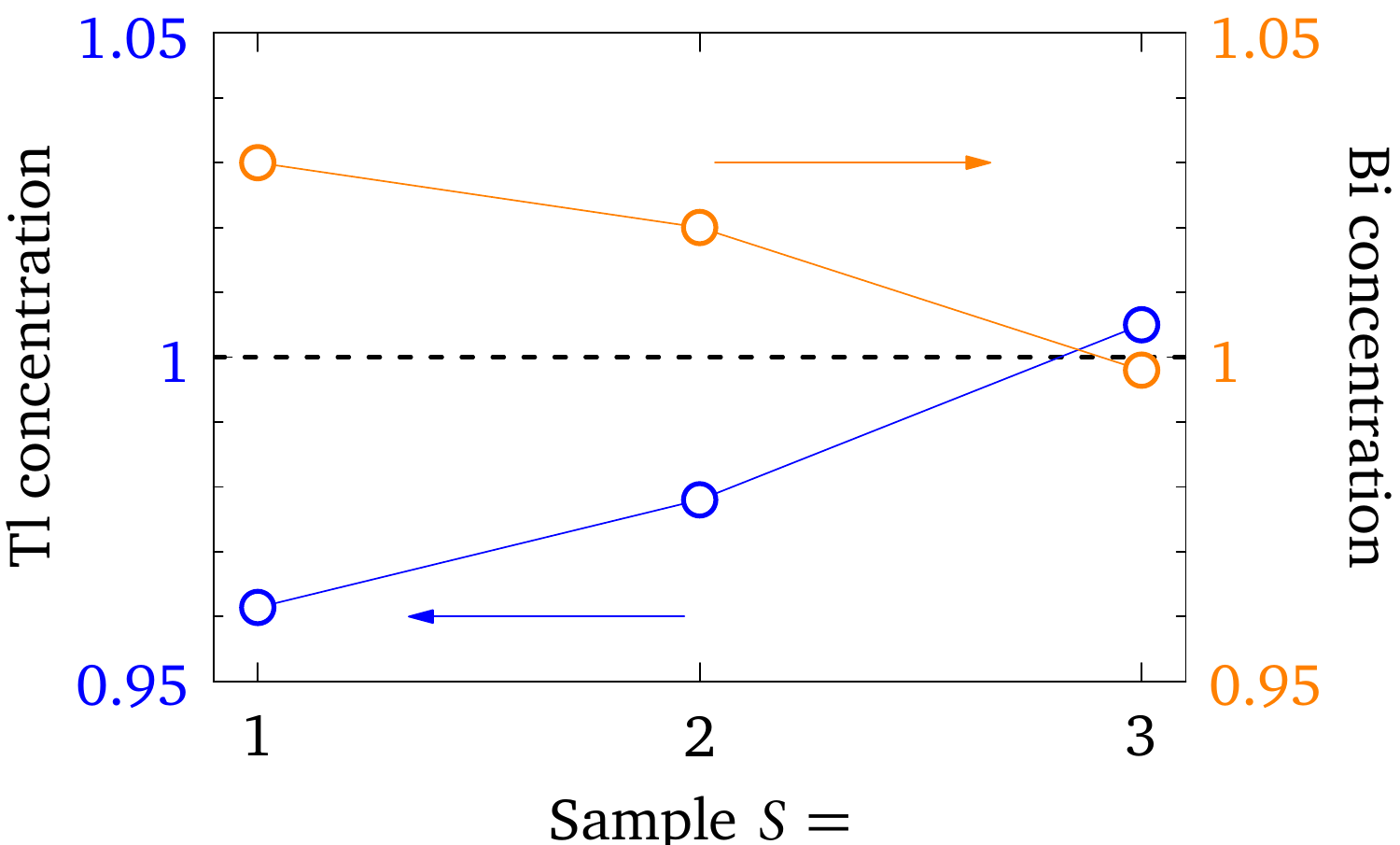}
	\caption{Evolution of the real content of Tl:Bi depending on the initial content of precursor in the synthesis.
		}
	\label{fig1_content}
\end{figure*}

We performed energy dispersive x-ray analysis (EDX) on our three samples $S_1$, $S_2$, and $S_3$ at an incident energy of 27 keV.
We extracted their real content ratio by assuming that the content of Tl + Bi = 2, as you can see on table \ref{TableEDX}.
\begin{table}[ht]
	\centering
	\caption{Elements content from EDX}
	\setlength{\tabcolsep}{1.5em} 
    	\renewcommand{\arraystretch}{1.8}
\begin{tabular}{@{}l  r | r r r r @{}}
\toprule
  Sample& & Tl &Bi&Se&S \\
\hline 
  $S_1$&\%&23.05&24.9&27.03&25.02 \\
  &ratio&0.96&1.04&1.12&0.96\\
\hline
  $S_2$&\%&23.32&24.34&26.46&25.88 \\
  &ratio&0.97&1.03&1.11&0.98\\
\hline
  $S_3$&\%&23.82&23.75&27.68&24.75 \\
  &ratio&1.01&0.99&1.05&1.04\\
\hline
\end{tabular}
\label{TableEDX}
\end{table}
This confirmed the presence of a small excess of Bi compared to Tl content on sample $S_1$ leading to the electron doping on the system. Including a higher nominal content of Tl in the synthesis decreases the Tl:Bi ratio and allows to obtain single crystals with an almost perfect stoichiometry, as seen in sample $S_3$ of Fig.~\ref{fig1_content}.
\subsubsection{XRD}
We obtained the lattice parameters in a hexagonal structure for our three samples $S_1$, $S_2$, and $S_3$ using single crystal x-ray diffraction as shown in table \ref{TableLat}. 
\textit{c} is almost constant and \textit{a} lightly decreases as the content of Tl:Bi approaches 1. 
The ratio \textit{c}/\textit{a} is an important parameter in order to get the system at the charge neutrality point since the gap of the Dirac cone in the electronic band structure is linked to this parameter \cite{Singh2012}. For the sample $S_3$, we obtained a ratio of around 5.322 which is in good agreement with the critical point predicted theoretically.
\begin{table}[ht]
	\centering
	\caption{Lattice parameters in the hexagonal structure for the three samples, $S_1$, $S_2$, and $S_3$, obtained by single crystal x-ray diffraction.} 
    	\renewcommand{\arraystretch}{1.5}
\begin{tabular}{@{}lrrr@{}}
\hline
\hline
& a ($\si{\angstrom}$)\hspace{1cm} & c ($\si{\angstrom}$)\hspace{1cm} &ratio (c/a)\\
\hline
\textbf{$S_1$}\hspace{2cm}&4.185\hspace{1cm} & 22.2\hspace{1cm} & 5.304\\

\hline
\textbf{$S_2$}\hspace{2cm}& 4.183\hspace{1cm} & 22.182\hspace{1cm} &5.302 \\

\hline
\textbf{$S_3$}\hspace{2cm} &4.173\hspace{1cm} & 22.212\hspace{1cm} & 5.322 \\
\hline
\end{tabular}
\label{TableLat}
\end{table}
\subsection{Band structure calculations} 
The electronic structure was calculated with the Quantum Espresso package \cite{Giannozzi2009}, employing Projector-Augmented Wave pseudopotentials to simulate the core interactions \cite{Kresse1999}. The PBE functional \cite{Perdew1996} was selected to reproduce the exchange-correlation effects, including the spin-orbit coupling (SOC) effect, as done in previous works \cite{Singh2012}. A converged plane wave basis set was used, with a kinetic energy cutoff at 50 Ry and 500 Ry for the wavefunctions and the electronic density, respectively. For the reciprocal space sampling we used a $8\times8\times8$ $k$-point grid with no shift with respect to the $\Gamma$ point, and the structural relaxations were performed until all the components of the forces were smaller than the convergence threshold $10^{-5}$ (a.u.), and the energy differences were smaller than $10^{-6}$ (a.u.). The unit cell parameters used as starting point for the structural optimization were taken from previous experimental studies \cite{YangXU2011}.
\begin{figure*}[!th]
	\includegraphics[width=0.6\linewidth]{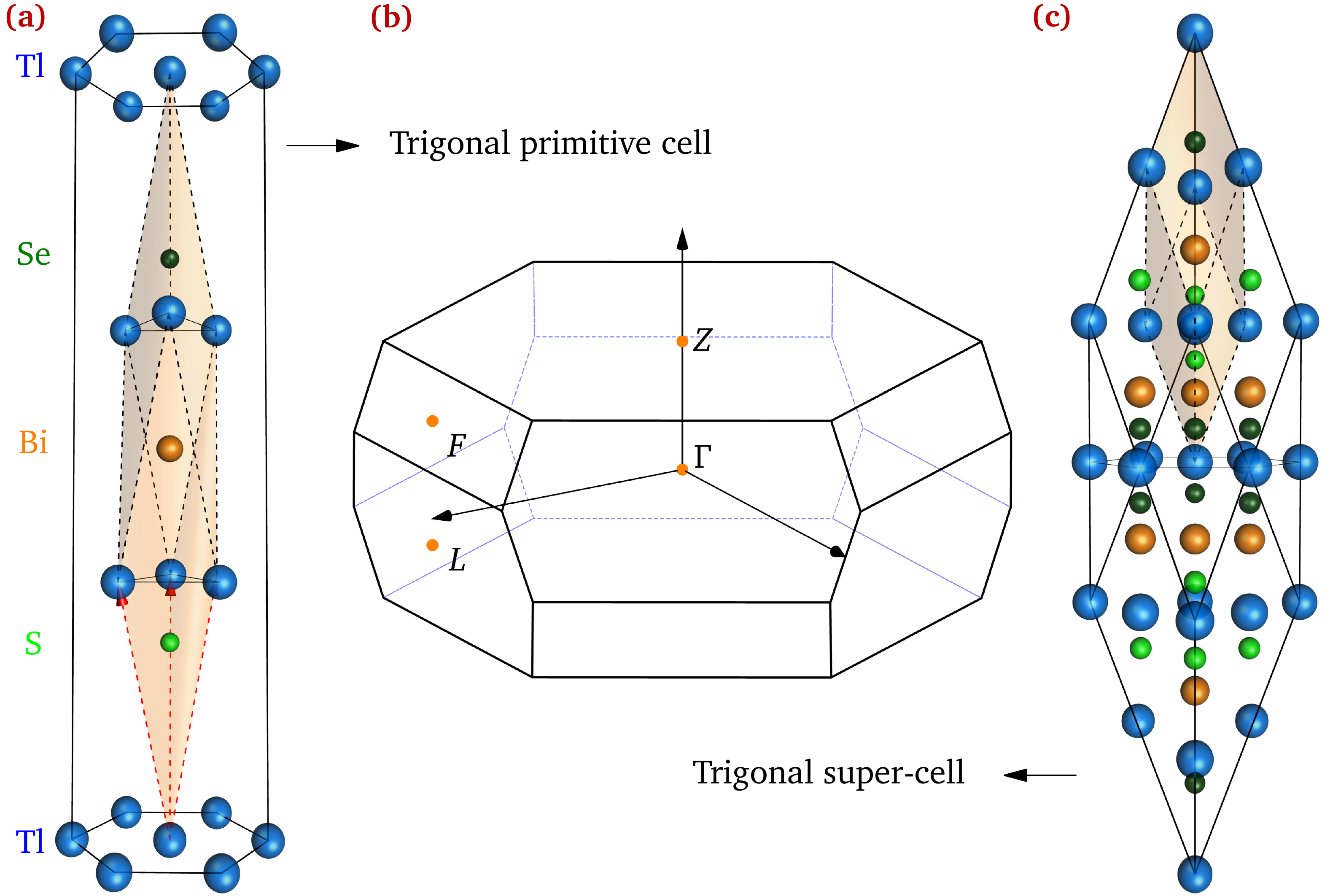}
	\caption{(a) Trigonal primitive unit cell of TlBiSSe with no inversion symmetry. (b) Sketch of the trigonal Brillouin zone of the primitive unit cell and (c)  Trigonal super-cell which contains $2\times2\times2$ trigonal primitive cells of Fig.~\ref{BZs}(a).}
	\label{BZs}
\end{figure*}
As observed experimentally, the TlBiSSe compound shows a doubly degenerated Dirac cone around the Gamma point, so the Kramers degeneracy is supposed to be present and inversion symmetry must be preserved \cite{Singh2012, Niu2012, Sato2011}. The inversion symmetry is preserved because the S and Se layers are not ordered, leading to a symmetric situation in the +k and -k direction if we take the average \cite{Singh2012}. However, the TlBiSSe trigonal primitive unit cell does not preserve this symmetry Fig.~\ref{BZs}(a), lifting the Kramers degeneracy when the soc effects are taken into account. Thus, we performed the calculations using a super-cell which contains $2\times2\times2$ trigonal primitive cells, see Fig.~\ref{BZs}(c), and exchanged the order of some of the S and Se layers in order to preserve the inversion symmetry. 
\begin{figure*}[!th]
	\includegraphics[width=0.95\linewidth]{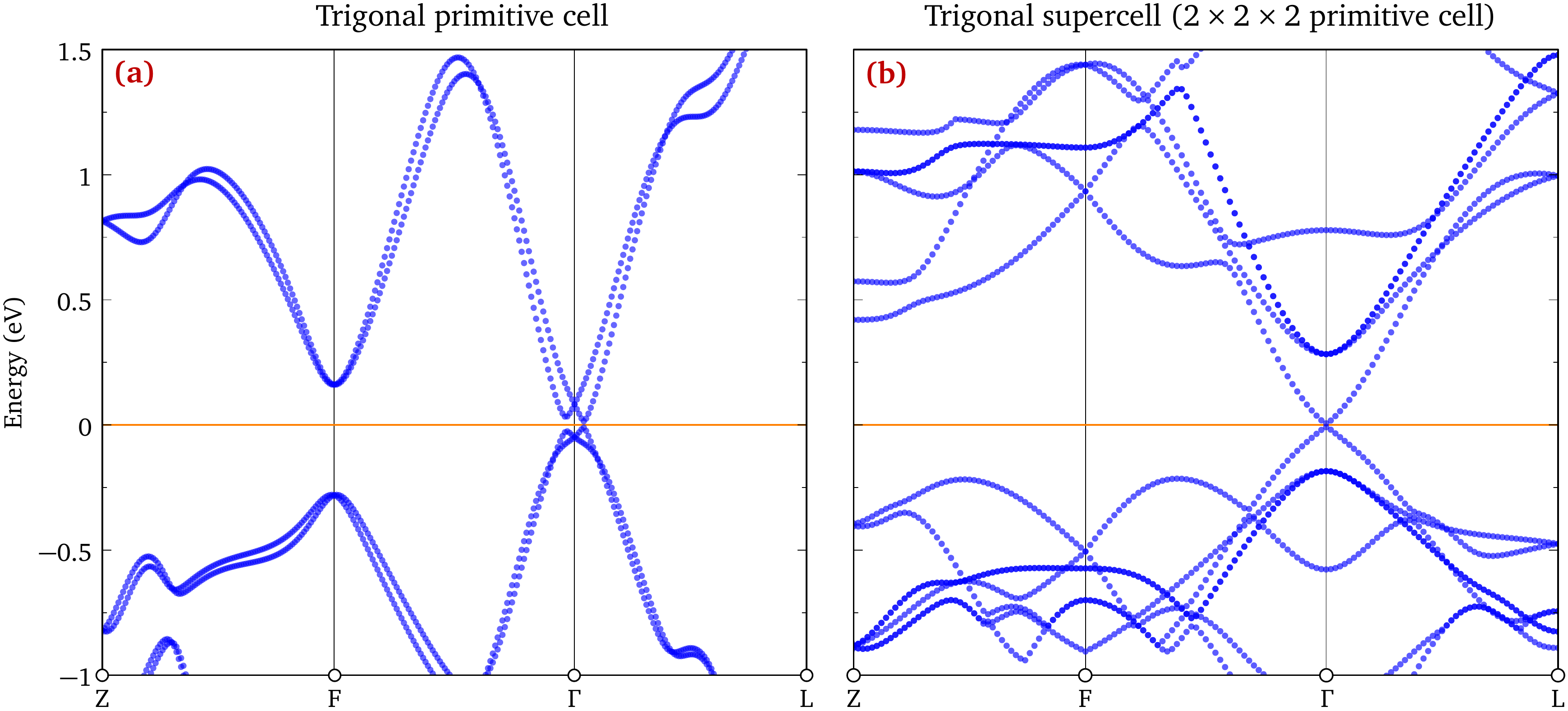}
	\caption{Calculated band structure of TlBiSSe in (a) the trigonal primitive cell and (b) in a super-cell which contains $2\times2\times2$ trigonal primitive cells with exchanged orders of some of the S and Se layers in order to preserve the inversion symmetry.}
	\label{bandStructure2}
\end{figure*}
In a first approximation, we performed the primitive cell calculation, in which the Kramers degeneracy is lifted, leading to a band structure corresponding to a Weyl semimetal, see Fig.~\ref{bandStructure2}(a). After that, the band structure of the super-cell with inversion symmetry, shown in Figure \ref{bandStructure2}(b), was obtained. In this band structure, the Kramers degeneracy is recovered and a doubly degenerate 3D Dirac cone is obtained around the $\Gamma$ point. 
\begin{figure*}[!th]
	\includegraphics[width=0.75\linewidth]{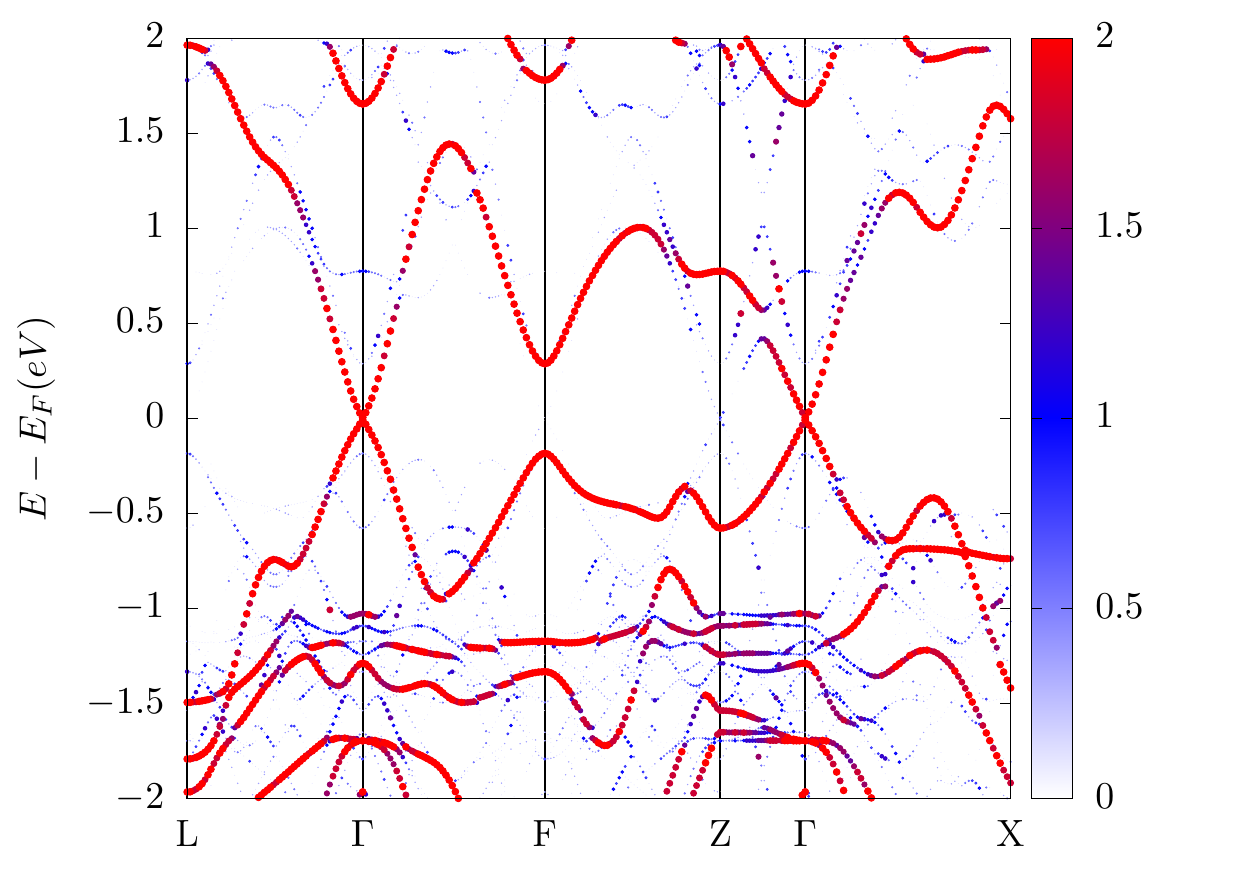}
	\caption{Calculated band structure of TlBiSSe in a trigonal primitive cell after unfolding the $2\times2\times2$ trigonal super-cell. The color scale represents the amplitude of the super-cell eigenstates projection into the primitive cell eigenstates.}
	\label{bandStructure3}
\end{figure*}
In order to recover an effective primitive cell band structure from the super-cell calculation, we performed an unfolding of the super-cell band structure using the BandUp code \cite{Medeiros2014, Medeiros2015}, shown in figure \ref{bandStructure3}. The relative weight of each point of the unfolded band structure was determined by the projection of the super-cell eigenstates into the primitive cell ones \cite{Popescu2012}, represented with different colors in figure \ref{bandStructure3}. We clearly see that this method conserved the inversion symmetry and the doubly degenerate Dirac cone arises in the primitive trigonal cell.
\subsection{Magneto-optical measurements} 
\begin{figure*}[!th]
	\includegraphics[width=0.98\linewidth]{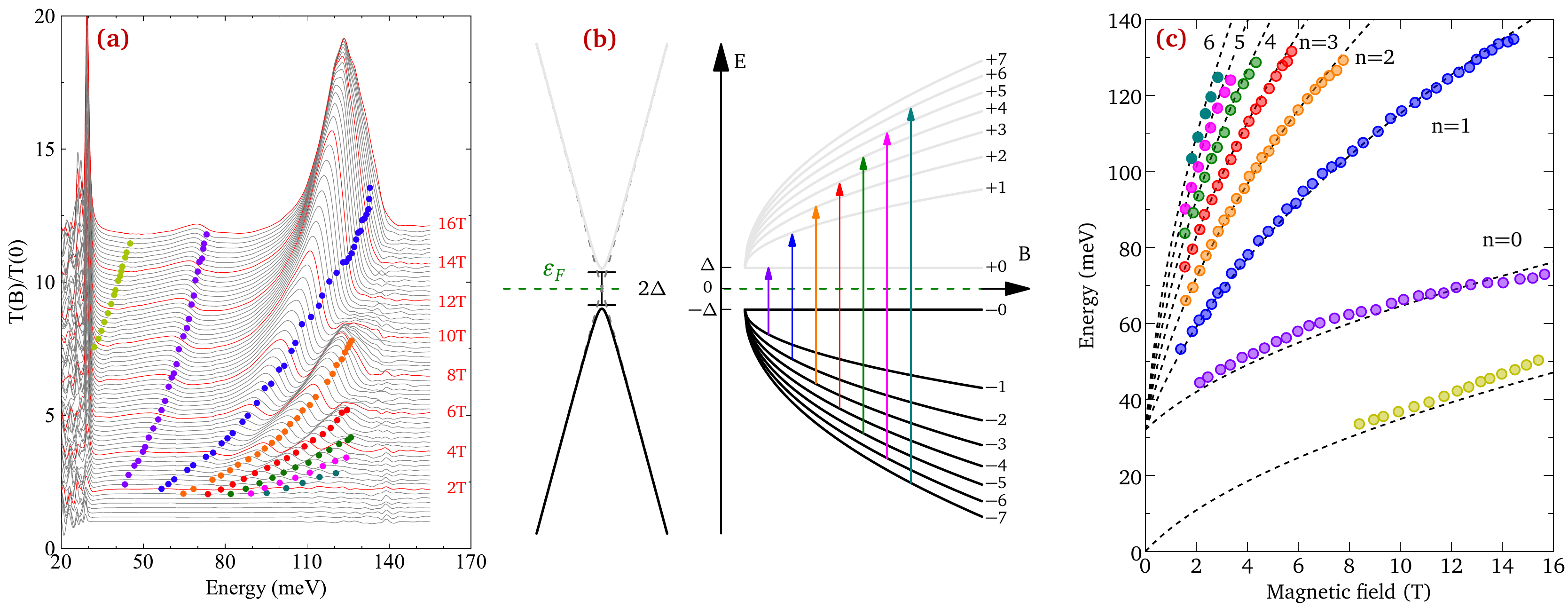}
	\caption{(a) Magneto-transmission normalized by zero-field transmitance, $T(B)/T(0)$, for B up to 16 T for TlBiSSe single crystal $S_3$. (b) Representation of gapped Dirac bands and LL dispersions as function of magnetic field. The arrows indicate one allowed possibility of inter-LL transitions across the gap. (c) Magnetic field dependence of the energy positions of the LLs fitted from Eq.~\ref{inter}. }
	\label{Mo_TR}
\end{figure*}

\begin{figure*}[!th]
	\includegraphics[width=0.7\linewidth]{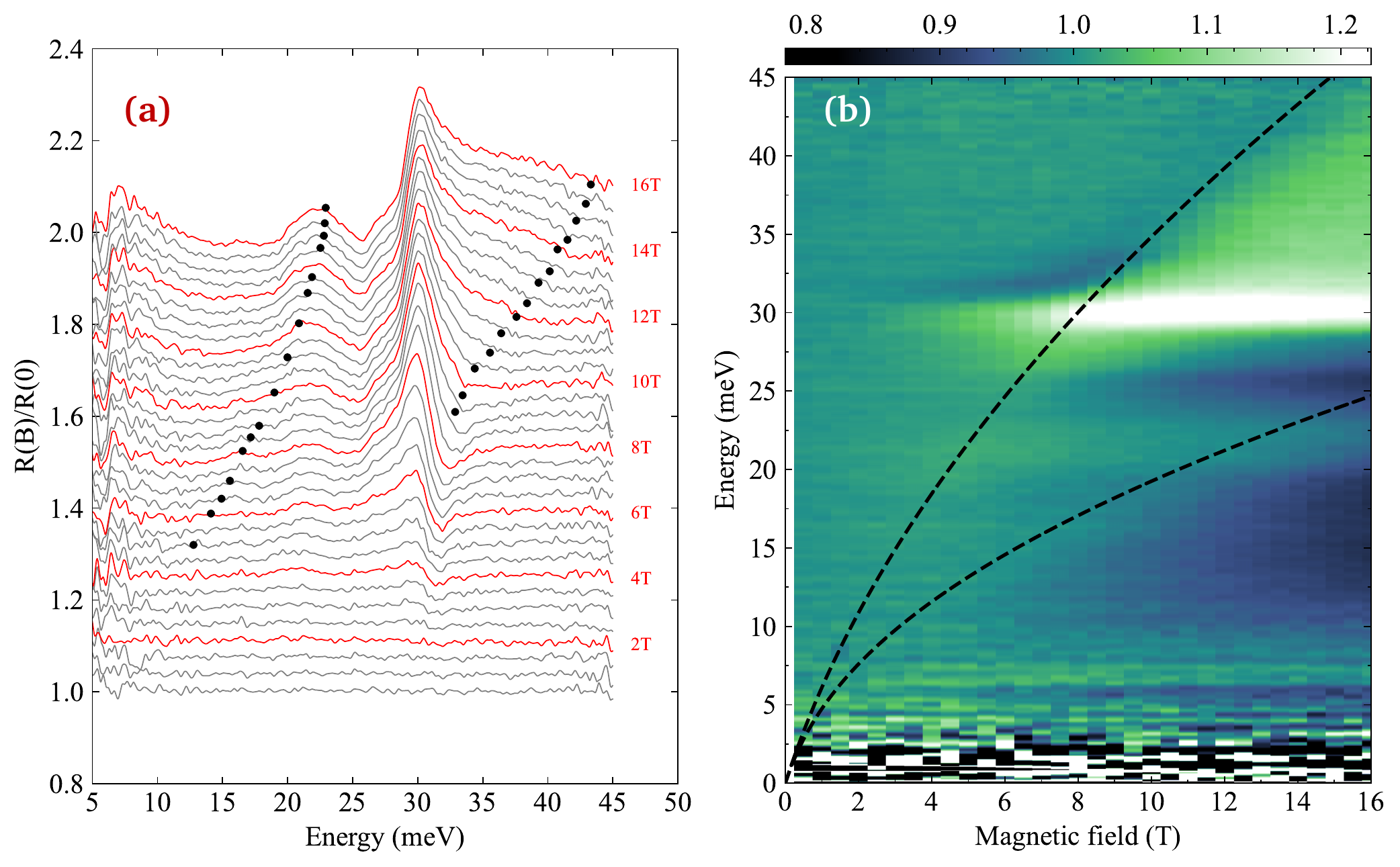}
	\caption{(a) Magneto-reflection normalized by zero-field reflectance $R(B)/R(0)$, for B up to 16 T of TlBiSSe single crystal $S_3$ at 4.2~K. (b) Color plot of the relative magneto-reflection, R$_B$/R$_0$, from Fig.~\ref{fig_MO}(a). Green dashed lines show the fit of the intra-LL excitations from transmission data, from Eq.~\ref{intra}.}
	\label{fig_MO}
\end{figure*}

%
We performed high magnetic field measurements in National High Magnetic Field Laboratory of Grenoble. Magneto-reflectance or transmittance was measured on sample $S_3$ using a superconducting coil up to 16 T. The sample was kept at 4.2~K in a low-pressure helium exchange gas. The magnetic field was applied in Faraday configuration, with the field direction parallel to the layer-stacking direction. 
Magneto-reflectance and magneto-transmittance data are shown in Fig.~\ref{fig_MO}. We extracted the positions of the inter- and intra-Landau level transitions looking for the maximum of R(B)/R(0) of the magneto-optical reflectance measurements or for a minimum of the first derivative of T(B)/T(0) of the magneto-optical Transmittance, see also the Fig.~\ref{Mo_TR}(a).
Figure \ref{Mo_TR}(b) shows a representation of gapped Dirac bands and their corresponding inter-LL excitations as a function of the magnetic field using Eq.~(\ref{inter}). The colored arrows indicate some of the allowed inter-LL transitions across the gap. The optical selection rule for TlBiSSe only allows Landau level transitions with $\Delta n = \pm 1$. Taking into account this selection rule and the gapped Dirac band dispersion shown in equation 1 from the main text, we obtained the  energy of the inter-LL and intra-LL transitions with the following equations:
\begin{equation}
E_n^{Inter} =\sqrt{2 \hbar e v_{x}v_{y} B \mid n+1 \mid + \Delta^{2}} + \sqrt{2 \hbar e v_{x}v_{y} B \mid n \mid + \Delta^{2}}.
\label{inter}
\end{equation}
\begin{equation}
E_n^{Intra} =\sqrt{2 \hbar e v_{x}v_{y} B \mid n+1 \mid + \Delta^{2}} - \sqrt{2 \hbar e v_{x}v_{y} B \mid n \mid + \Delta^{2}}.
\label{intra}
\end{equation}
With $\Delta$ the optical gap, and $v_{x}, v_{y}$ the two in-plane velocities of the Dirac cone.
\\
The relative magneto-reflectivity $R_B/R_0$ for a lower energy range, 3--45~meV is shown in Fig.~\ref{fig_MO}. The most obvious feature is a strong line showing up at 30~meV in fields above $\sim 3$~T. This line coincides with a weak zero-field feature, possibly a phonon line, as shown in Fig.~2(e) from the main text.
Another feature is a broad line at $\sim 23$~meV, which again corresponds to a weak phonon mode.
The two dashed green lines are intra-LL transitions, obtained from the above fit. 
It is possible that the background intensity is described by these intra-LL transitions, in particular above 8~T as the higher intra-LL transition crosses the 30~meV line.
\subsection{Electrical resistivity measurements}
\begin{figure}[htb]
	\includegraphics[width=10cm]{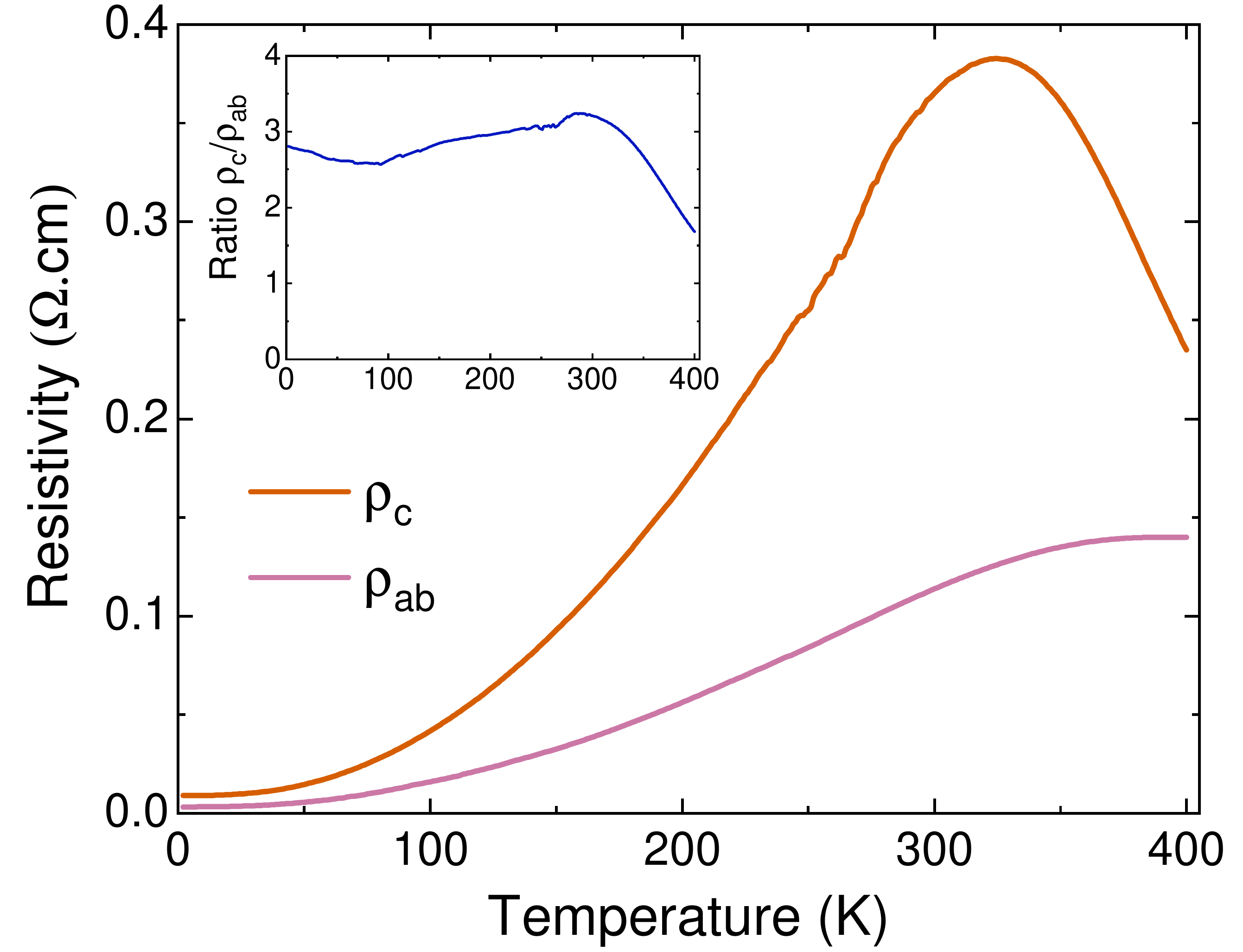}
	\caption{Resistivity transport measurements performed on sample S$_3$ as a function of temperature in- and out- of plane direction, $\varrho_{xy}$ and $\varrho_z$ respectively. The inset show the temperature dependence resistivity ratio $R = \frac{\varrho_z}{\varrho_{{xy}}}$.
	}
	\label{fig_transport}
\end{figure}

We performed electrical resistivity measurements in the \textit{ab}-plane (or \textit{xy}-plane) and along the \textit{c}-axis (\textit{z} direction) as a function of temperature from 400K to 1.8K as shown on Fig.~\ref{fig_transport}, in order to characterize the anisotropy in the system. Electrical contacts were made using silver paint in Van der Pauw configuration and using a ring array for the resistivity on the ab-plane and along the c-axis, respectively. Measurements were performed by a Physical Properties Measurement System (PPMS; Quantum Design). 
Both the ab-plane and c-axis resistivity measurements show a bad metallic behavior with a RRR of around 30 indicating a relatively low impurity concentration in the system. The resistivity values at 1.8K are  $\rho_{ab}$(1.8K) = 3.2 m$\Omega$ cm and $\rho_{c}$ (1.8K) = 9 m$\Omega$ cm. Our in-plane resistivity measurements agree well with the previous study \cite{Novak2015}. We extract an anisotropic resistivity ratio R at T $\approx$ 0~K:
\begin{equation}\label{b3}
 R = \frac{\varrho_z}{\varrho_{{xy}}} \approx 2.7.
\end{equation}
\subsection{Extraction of the anisotropic Dirac cone parameters} 
%
%
In this section, we will discuss how to extract the three velocities ($v_x, v_y, v_z$)  that describe the slope of the Dirac cone in sample $S_3$ in the presence of an anisotropy in the linear dispersion. For this purpose, we will use the DC anisotropy resistivity ratio, and optical and magneto-optical measurements performed on sample $S_3$.
The real part of the optical conductivity of an anisotropic 3D Dirac system \cite{Rukelj2020,Lim_2020,Kotov2016} for incoming photon frequency $\omega$ and  with an electric field 
pointing in the $x$-direction
\begin{equation}
{\rm{Re}} \, \sigma_{x}(\omega) = g \frac{\sigma_0}{6\pi \hbar} \frac{v_x}{v_yv_z} \hbar \omega \, \Theta(\hbar\omega - 2\varepsilon_F).
\end{equation}
Similarly, for the electric field pointing in the  $y$-direction the optical conductivity is
\begin{equation}
{\rm{Re}} \, \sigma_{y}(\omega) = g \frac{\sigma_0}{6\pi \hbar} \frac{v_y}{v_xv_z} \hbar \omega \, \Theta(\hbar\omega - 2\varepsilon_F).
\end{equation}
The real part of the total in-plane ($xy$-plane) optical conductivity is 
\begin{equation}\label{b111}
\sigma_1(\omega) = {\rm{Re}} \, \sigma(\omega) = \frac{1}{2} \left( {\rm{Re}} \, \sigma_{x}(\omega) + {\rm{Re}} \, \sigma_{y}(\omega) \right) = g \frac{\sigma_0}{6\pi \hbar} \frac{v_x^2 + v_y^2}{2v_x v_y v_z} \hbar \omega \, \Theta(\hbar\omega - 2\varepsilon_F)
\end{equation}
If we take into account such a non-isotropic linear dispersion into our zero field reflection measurements, performed with an in-plane polarized light, then $v_F$ of sample $S_3$ is given by:
\begin{equation}\label{b2}
    \frac{1}{v_F} = \frac{v_x^2 + v_y^2}{2v_x v_y v_z}  \approx \frac{1}{1.8\times 10^5 \, \text{ m/s}}
\end{equation}
From the zero field optical measurements we have \eqref{b2}, and from our the magneto optical measurements and the fit using \eqref{intra}, we get:
\begin{equation}\label{b1}
\sqrt{v_xv_y} \approx 4\times10^5 \, \text{ m/s}.
\end{equation}
In an anisotropic 3D Dirac system, the Drude concentration of the charge carriers density can be expressed like as \cite{Rukelj2020}:
\begin{equation} \label{a44}
 n_{\al} =\frac{g}{V}\sum_\kk m_e v_\al^2(\kk) \delta(\e_F - \e(\kk))
\end{equation}
Where $m_e$ is the bare electron mass and the degeneracy factor $g=2$. Moreover, using the energy dispersion in this general case $\e(\kk) = \sqrt{\sum_i (\h v_i k_i)^2 }$, we get:
\begin{eqnarray}\label{a55}
 && n_{x} = \frac{1}{3} \frac{v_x}{v_y v_z} \frac{m_e}{\pi^2 \h^3} \e_F^2, \nonumber \\
 && n_{y} = \frac{1}{3} \frac{v_y}{v_x v_z} \frac{m_e}{\pi^2 \h^3} \e_F^2,  \nonumber \\
 && n_{z} = \frac{1}{3} \frac{v_z}{v_x v_y} \frac{m_e}{\pi^2 \h^3} \e_F^2,
\end{eqnarray}
where $\varepsilon_F$ is the Fermi energy. Starting from the general expression of the Drude DC term $\sigma^{DC}_{\alpha} \sim e^2 n_{\alpha}/m_e $  \cite{Rukelj2020} and using \eqref{a55} with  $ n_{xy} = (n_{x} + n_{y})/2$, we obtain an expression for the resistivity ratio:
\begin{equation}\label{b33}
 R = \frac{\sigma^{DC}_{{xy}}}{\sigma^{DC}_z} = \frac{n_{xy}}{n_z} = \frac{n_{x} + n_{y}}{2n_z} = \frac{v_x^2 + v_y^2}{2v_z^2} \approx 2.7.
\end{equation}
Dividing \eqref{b33} with \eqref{b2} we obtain: 
\begin{equation}\label{cucu}
 \frac{v_x v_y}{v_z} \approx 4.86\times 10^5 \, \text{ m/s},
\end{equation}
and by using \eqref{b1}, we finally get:
\begin{equation}\label{vz}
 v_z \approx  3.3\times 10^5 \, \text{ m/s}.
\end{equation}
If \eqref{vz} is inserted back into \eqref{b2} we get the following expression which can be further shaped
\begin{eqnarray}\label{pri}
&& \frac{v_xv_y}{v_x^2 + v_y^2} = \frac{1}{3.66} \to v_x^2 + v_y^2 = 3.66 v_xv_y, \nonumber \\
&& \to (v_x -v_y)^2 = 1.66v_xv_y \to v_x -v_y = \sqrt{v_xv_y} \sqrt{1.66} \nonumber \\
&& v_x -v_y \approx 5.16\times 10^5 \, \text{ m/s}. \nonumber \\
\end{eqnarray}
Together \eqref{b1} and \eqref{pri} give a quadratic equation whose solutions are 
\begin{equation}
 v_x \approx 7.34\times 10^5 \, \text{ m/s}, \hspace{3mm}  v_y \approx 2.18\times 10^5 \,\text{ m/s}
\end{equation}
A general procedure can be developed for determining the band parameters. Let's go back to \eqref{b1}, \eqref{b2} and \eqref{b33} and introduce the following constants
\begin{equation}\label{bgot}
  \frac{v_x^2 + v_y^2}{v_x v_y v_z} = \frac{1}{v_1}, \hspace{3mm} \sqrt{v_xv_y}= v_2, \hspace{3mm} \frac{v_x^2 + v_y^2}{v_z^2}= 2R.
\end{equation}
Then determining $v_z$ following the mentioned recipe from above 
\begin{equation}\label{ssw}
 v_z = \frac{v_2^2}{2Rv_1},
\end{equation}
Inserting \eqref{ssw} into the first term in \eqref{bgot} we get 
\begin{equation}
 v_x^2 + v_y^2 = \frac{1}{2R} \frac{v_2^2}{v_1^2} v_xv_y,
\end{equation}
which has a solution only if 
\begin{equation}\label{uvijet}
 \frac{v_2^2}{v_1^2} > 4R.
\end{equation}
If we had initially started with a double degenerate Dirac cone with $g=4$ in the 3D conical band formula \eqref{b111}, the fitting procedure would have given $v_1 = 1.8\times10^5$m/s and the inequality \eqref{uvijet} would be violated. This means that there are no $v_y$ and $v_x$ that satisfy the set of equations.
But if we take $g=2$, for a spin degenerated Dirac cone, as we did in fitting the optical conductivity, we get $v_1 = 0.9\times 10^5$m/s and the above inequality is fulfilled.
In this approach we can write down the general expressions for $v_x$ and $v_y$ once the condition \eqref{uvijet} holds:
\begin{eqnarray}
 && v_y = \frac{v_2}{2} \left( \sqrt{\frac{1}{2R}\frac{v_2^2}{v_1^2} + 2} \, - \sqrt{\frac{1}{2R}\frac{v_2^2}{v_1^2} - 2} \,  \right), \nonumber \\
 && v_x = \frac{v_2}{2} \left( \sqrt{\frac{1}{2R}\frac{v_2^2}{v_1^2} + 2} \, + \sqrt{\frac{1}{2R}\frac{v_2^2}{v_1^2} - 2} \,  \right).
\end{eqnarray}
Multiplying the above velocities to check that $v_xv_y = v^2_2$ and using \eqref{ssw} we obtain
\begin{equation}
 v_{x,y} = \frac{v_2}{2}\left( \sqrt{\frac{v_z}{v_1} + 2} \, \pm \sqrt{\frac{v_z}{v_1} - 2} \,  \right).
\end{equation}
\subsection{Determination of the Fermi level on sample $S_1$ and $S_2$} 
Finally, we want to determine the Fermi energy $\e_F$ for our two doped samples $S_1$ and $S_2$ using the zero field reflectivity measurements and the 
rigid band approximation.
We turn our attention to the real part of the low-energy in $xy$-plane dynamical dielectric tensor 
\begin{equation} \label{a2}
 {\rm{Re}} \, \epsilon(\om) = \epsilon_\infty - \om_{pl}^2/\om^2
\end{equation}
where $\epsilon_\infty$ is the $ {\rm{Re}} \, \epsilon(\om = 0)$ values originating from all electronic interband excitations and the infrared active phonon modes. In \eqref{a2} $\h^2\om^2_{pl} = \h^2 e^2n_{xy}/(m_e\epsilon_0 \epsilon_\infty)$ is the screened plasma energy squared and  $m_e$ is the bare electron mass.
From the definition of the plasma energy, we get using $ n_{xy} = (n_{x} + n_{y})/2$ and \eqref{a55}:
\begin{equation}  \label{a6}
\h^2\om^2_{pl} = \frac{ e^2}{\epsilon_0 \epsilon_\infty} \, \frac{1}{6} \frac{v_x^2 + v_y^2}{v_x v_y v_z} \,  \frac{\e_F^2 }{\pi^2 \h}, 
\end{equation}
From the fit of the reflectivity measurements we extracted for the two doped samples and the values of $\om_{pl}$ and $\epsilon_\infty$, we obtained the following Fermi levels using Eqs.~\eqref{a6} and \eqref{b2}:\\
\begin{table}[ht]
	\centering
	\setlength{\tabcolsep}{1.5em} 
    	\renewcommand{\arraystretch}{1.8}
\begin{tabular}{@{}l | r r r @{}}
\toprule
  Sample & $\h\om_{pl}$ (eV) & $\epsilon_\infty$ & $\e_F$ (eV) \\
\hline 
  $S_1$& 0.123 & 28.4 & 0.293 \\
\hline
  $S_2$& 0.107 & 24.6 & 0.237 \\
\hline
\end{tabular}
\label{onset}
\end{table}
\\%
This sets the Pauli edge for sample $S_1$ and $S_2$ at $2\e_F = 0.586$~eV and $0.474$~eV, respectively. Since clearly $\e_F \gg \Delta$ even if we had a massive 3D Dirac, we are at high enough Fermi energies, that the gap has no influence on the DC zero field transport properties.
We can also determine the total concentration of electrons in the Dirac cone to be $n_1= 5.55 \times 10^{19}$ cm$^{-3}$ and $n_2= 2.94 \times 10^{19}$ cm$^{-3}$ for sample $S_1$ and $S_2$ respectively using the formula:
\begin{equation}
 n = \frac{ \e_F^3}{3\pi^2 \h^3 v_x v_y v_z}.
\end{equation}
\\
\twocolumngrid
\bibliography{TlBiSSe}

\end{document}